\documentclass{PoS}

\newcommand{\SU}{\mathrm{SU}}
\newcommand{\Lren}{L_{\mbox{\tiny{ren}}}}
\newcommand{\Real}{\mathrm{Re}}
\newcommand{\tr}{\mathrm{tr}}
\newcommand{\eq}[1]{\begin{equation}\label{#1}}
\newcommand{\en}{\end{equation}}
\newcommand{\ear}[1]{\begin{eqnarray}\label{#1}}
\newcommand{\enar}{\end{eqnarray}}

\title{Renormalization of Polyakov loops in different representations and the large-$N$ limit}

\ShortTitle{Renormalization of Polyakov loops in different representations and the large-$N$ limit}

\author{Anne Mykkanen, \speaker{Marco Panero} and Kari Rummukainen\\%
        Department of Physics and Helsinki Institute of Physics, University of Helsinki\\
        FIN-00014 Helsinki, Finland\\
        E-mail: \email{anne-mari.mykkanen@helsinki.fi}, \email{marco.panero@helsinki.fi}, \email{kari.rummukainen@helsinki.fi}
}

\abstract{We study the renormalization of Polyakov loops in different irreducible representations of $\SU(N)$ Yang-Mills theories at finite temperature, and investigate their behavior in the large-$N$ limit.
\vspace{4cm}
\begin{flushright}
HIP-2011-15/TH
\end{flushright}}

\FullConference{XXIX International Symposium on Lattice Field Theory \\
		 July 10 -- 16 2011\\
		 Squaw Valley, Lake Tahoe, California}

\begin{document}

\section{Introduction and motivation}
\label{sec:intro}

In Yang-Mills theories at finite temperature $T$, the deconfinement transition is described in terms of the spontaneous breakdown of an \emph{exact} center symmetry: the associated order parameter is the trace of the Polyakov loop $L$, which describes a static probe color charge in the thermal medium. However, the free energy associated with the \emph{bare} Polyakov loop is a divergent quantity in the continuum limit, and hence requires renormalization~\cite{Dotsenko:1979wb}. In this contribution, we discuss our ongoing study of renormalized Polyakov loops in $\SU(N)$ Yang-Mills theories with a different number of colors $N$, and in different irreducible representations of the gauge group.

There are many reasons for studying non-Abelian gauge theories with a large number of colors. The large-$N$ limit at fixed 't~Hooft coupling $\lambda=g^2 N$ and fixed number of flavors $N_f$ clarifies some non-trivial features of QCD~\cite{'tHooftlargeN}, and leads to a topological classification of Feynman diagrams. These can be arranged in expansions in powers of $1/N$, which has analogies with similar expansions in closed string theory~\cite{Aharony:1999ti}. The aspects related to volume reduction at large $N$~\cite{volume_reduction} have been reviewed in the plenary talk by \"Unsal~\cite{Unsal_proceedings}. For the phase diagram of QCD-like theories, the large-$N$ limit also has interesting implications for new phases at high density~\cite{McLerran:2007qj}. Finally, the large-$N$ limit also plays a technically crucial r\^ole in studies of the strongly interacting plasma~\cite{Gubser:2009md} based on the conjectured gauge/string correspondence~\cite{Maldacena_conjecture}.

The relevance of the large-$N$ limit for the physical case of three colors is an issue which can be tested in a first-principle approach, via lattice computations, and indeed this has been done with success in previous studies---see ref.~\cite{Teper:2009uf} for a review. In particular, as it concerns the finite-temperature properties, there is now convincing lattice evidence for the relevance of the large-$N$ limit for the Yang-Mills equation of state, both in $D=3+1$ and in $D=2+1$ dimensions~\cite{largeNpressure}. One may then wonder, whether this holds for other thermal quantities, too. From this point of view, a particularly interesting observable is the renormalized Polyakov loop $\Lren$: for this quantity, perturbative expansions around the high-temperature limit predict a non-monotonic behavior as a function of the temperature~\cite{perturbative_Polyakov_loop}, while holographic arguments suggest positive $\partial \Lren / \partial T$, instead~\cite{Noronha}. Moreover, it has also been pointed out that existing lattice results for the logarithm of $\Lren$ in $\SU(3)$ Yang-Mills theory reveal a characteristic $T^{-2}$ dependence~\cite{Megias_Ruiz_Arriola_Salcedo}.

Looking at the Polyakov loop in different irreducible representations allows one to test the Casimir scaling of the associated free energies~\cite{Casimir_scaling_tests}, and to investigate the  equivalence of different irreducible representations in the large-$N$ limit, and the possible implications for certain effective models for the physics of the deconfined plasma close to the deconfinement temperature $T_c$~\cite{Pisarski:2000eq}. Moreover, the finite-$T$ properties of strongly coupled gauge theories with \emph{dynamical} fermions in different representations are also interesting for Extended Technicolor models~\cite{ETC}.

\section{Setup of the computation}
\label{sec:computation_setup}

We study $\SU(N)$ Yang-Mills theories with $2 \le N \le 8$ colors on isotropic hypercubic lattices, using both the standard Wilson gauge action~\cite{Wilson:1974sk}:
\begin{equation}
S = \beta \sum_{x} \sum_{\mu < \nu} \left\{ 1 - \frac{1}{N}\Real~\tr U^{1,1}_{\mu,\nu}(x) \right\} 
\end{equation}
and the tree-level Symanzik-improved action~\cite{improved}:
\begin{equation}
\label{eq:improved_action}
S = \beta \sum_{x} \sum_{\mu < \nu} \left\{ 1 - \frac{1}{N}\Real~\tr \left[ \frac{5}{3} U^{1,1}_{\mu,\nu}(x) - \frac{1}{12} U^{1,2}_{\mu,\nu}(x) - \frac{1}{12} U^{1,2}_{\nu,\mu}(x) \right] \right\} 
\end{equation}
with $\beta=2N/g_0^2$, where $U^{1,1}_{\mu,\nu}(x)$ denotes the usual plaquette, while $U^{1,2}_{\mu,\nu}(x)$ denotes the ordered product of links around the rectangle with corners on the sites of coordinates $x$, $x+a\hat\mu$, $x+a\hat\mu+2a\hat\nu$ and $x+2a\hat\nu$. Our simulations are performed with an algorithm based on standard heat-bath and overrelaxation updates on $\SU(2)$ subgroups~\cite{algorithm}.

For the simulations with the Wilson action, highly accurate non-perturbative determinations of the physical scale are available in the literature~\cite{Wilson_scale}. To set the scale for our simulations with the improved action, we calculate the $T=0$ static potential $V(r)$ from expectation values of Wilson loops $\langle W(r,L) \rangle$:
\begin{equation}
\label{eq:plateau}
V(r) = a^{-1} \lim_{L\to \infty} \ln \frac{\langle W(r,L-a) \rangle}{\langle W(r,L) \rangle} 
\end{equation}
using five levels of smearing for the spacelike links. The values of $V(r)$ thus obtained are then fitted to the Cornell potential:
\begin{equation}
\label{eq:cornell}
V(r) = \sigma r + V_0 + \frac{\gamma}{r}
\end{equation}
to extract $\sigma$, $V_0$ and $\gamma$ (in appropriate units of the lattice spacing $a$). 

The computation of Polyakov loops in different irreducible representations can be done efficiently using the representation composition laws encoded in Young calculus. In particular, this leads to very simple relations for the $\SU(N)$ groups of lowest ranks: for $\SU(2)$, any irreducible representation $r$ can be labelled by a non-negative integer $n$ (such that dim $ r = n+1$, and the corresponding ``spin'' is $j=n/2$), and all characters are obtained from the fundamental one ($\tr_1 g$) via the recursive formula:
\begin{equation}
\tr_{n+1} g =\tr_{n} g~\tr_{1} g -\tr_{n-1} g \;\;\; \mbox{with:}\;\tr_{0} g=1 .
\end{equation}
For $\SU(3)$, the computation of characters for higher representations is greatly simplified by the following relation between the fundamental ($f$) and the two-index antisymmetric representation, which is nothing but the anti-fundamental ($\bar f$):
\begin{equation}
\frac{1}{2} [ ( \tr_f g )^2 - \tr_f ( g^2 ) ] = \tr_{~\bar f} g = ( \tr_f g )^\star .
\end{equation}
For $\SU(N>3)$, the traces in higher representations can be computed efficiently by combining Young calculus relations with the Weyl formula~\cite{Weyl_formula}:
\begin{equation}
\tr_{\vec\lambda} g = \frac{\det F(\vec{\lambda})}{\det F(\vec{0})} ,
\end{equation}
where the components of the $\vec \lambda$ vector are the lengths of the rows of the Young diagram of the representation, from top to bottom (with $\lambda_N=0$),
while $F_{kl}(\vec{\lambda}) = \exp\left[ i \left( N+\lambda_l-l \right) \alpha_k \right]$ and $e^{i \alpha_1}$, $e^{i \alpha_2}$, $\dots$ $e^{i \alpha_N}$ denote the eigenvalues of $g$ in the fundamental representation. In our code, the Polyakov loops in the twelve lowest non-trivial, independent irreducible representations for each gauge group are computed (the largest representation for the $\SU(8)$ gauge group that we computed is the $\mathbf{336}$). 

The relation between ``bare'' ($L$) and ``physical'' ($\Lren$) Polyakov loops in a given representation can be expressed by a multiplicative renormalization:
\begin{equation}
\langle L \rangle = Z^{-N_t} \langle \Lren \rangle ,
\end{equation}
where $N_t=1/(aT)$ denotes the number of lattice sites in the compactified Euclidean time direction. Note that $\langle L \rangle$ depends on both the bare coupling $g_0$ and on the temperature, while $\langle \Lren \rangle$ depends only on $T$, and the renormalization factor $Z$ depends only on $g_0$. If Casimir scaling holds, the $Z$'s for different representations are all related to the one for the fundamental representation.

In the literature, various methods to determine the renormalization factor $Z(g_0^2)$ have been proposed~\cite{renormalization_methods}; we chose to determine it from the constant term in the zero-temperature $Q\bar{Q}$ potential:
\begin{equation}
\label{eq:Z}
Z=\exp(aV_0/2) .
\end{equation}

\section{Preliminary results}
\label{sec:preliminary_results}

\begin{figure}[-t]
  \centerline{\includegraphics[width=.33\textwidth]{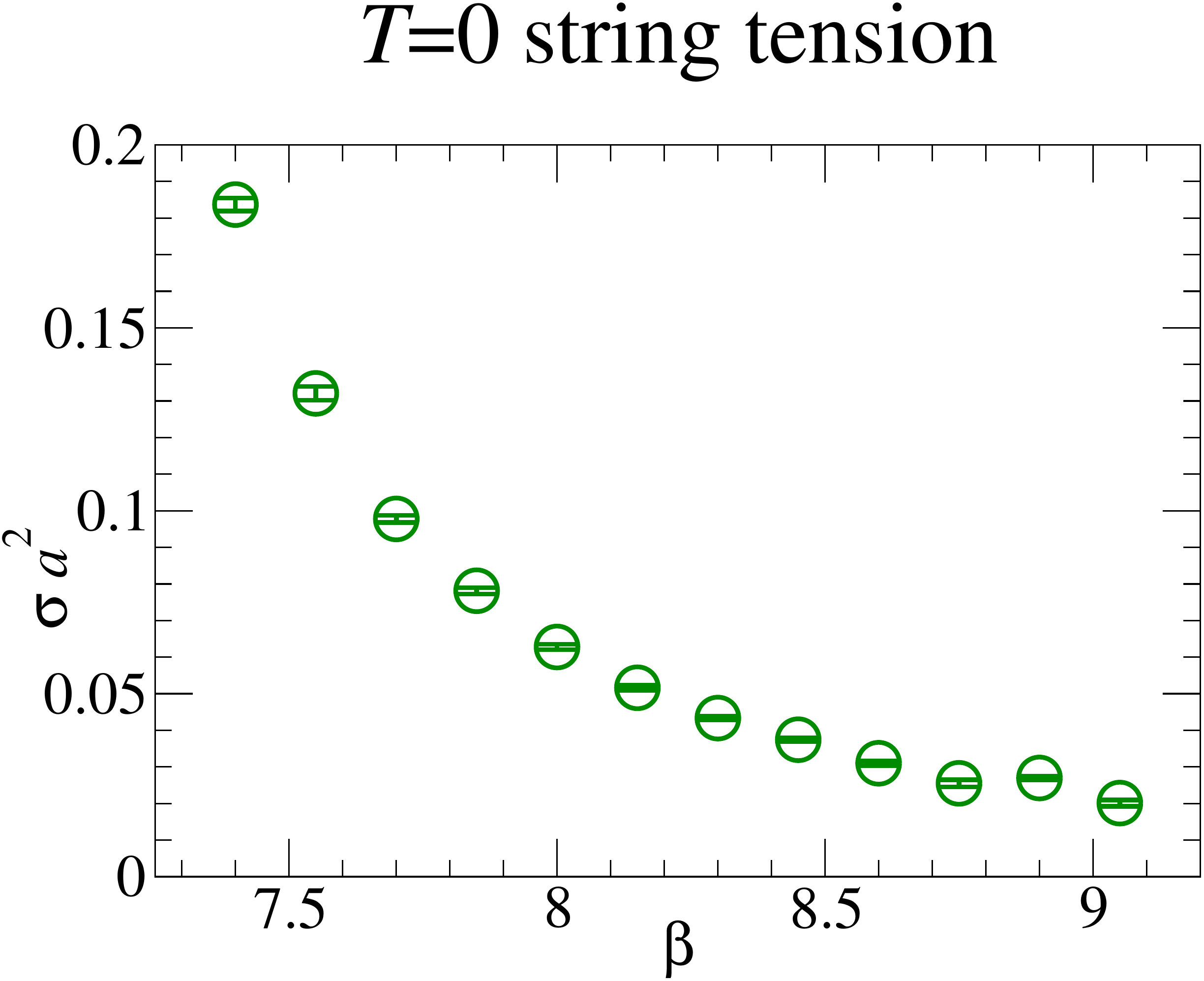} \hspace{4mm} \includegraphics[width=.317\textwidth]{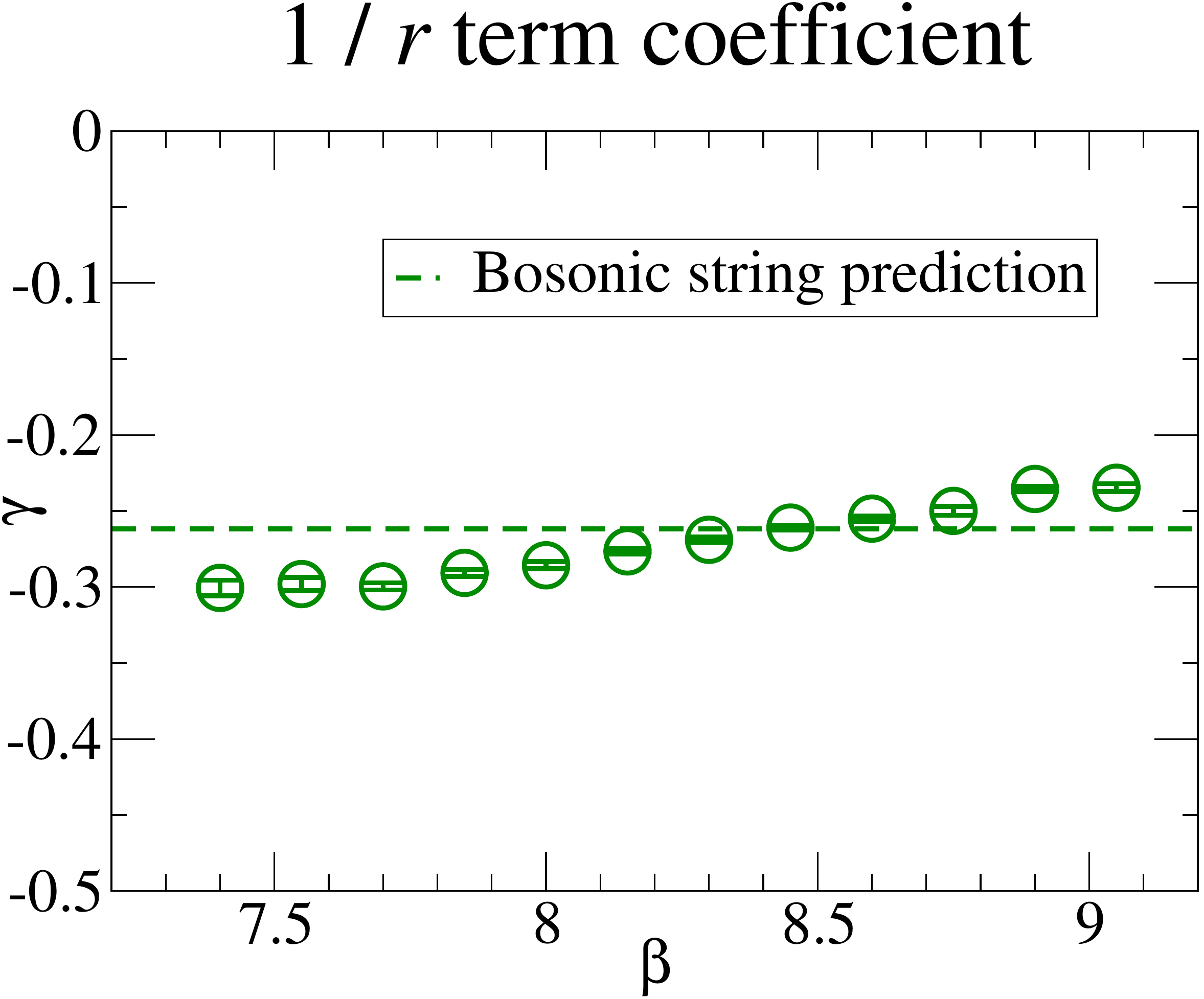} \hspace{4mm} \includegraphics[width=.332\textwidth]{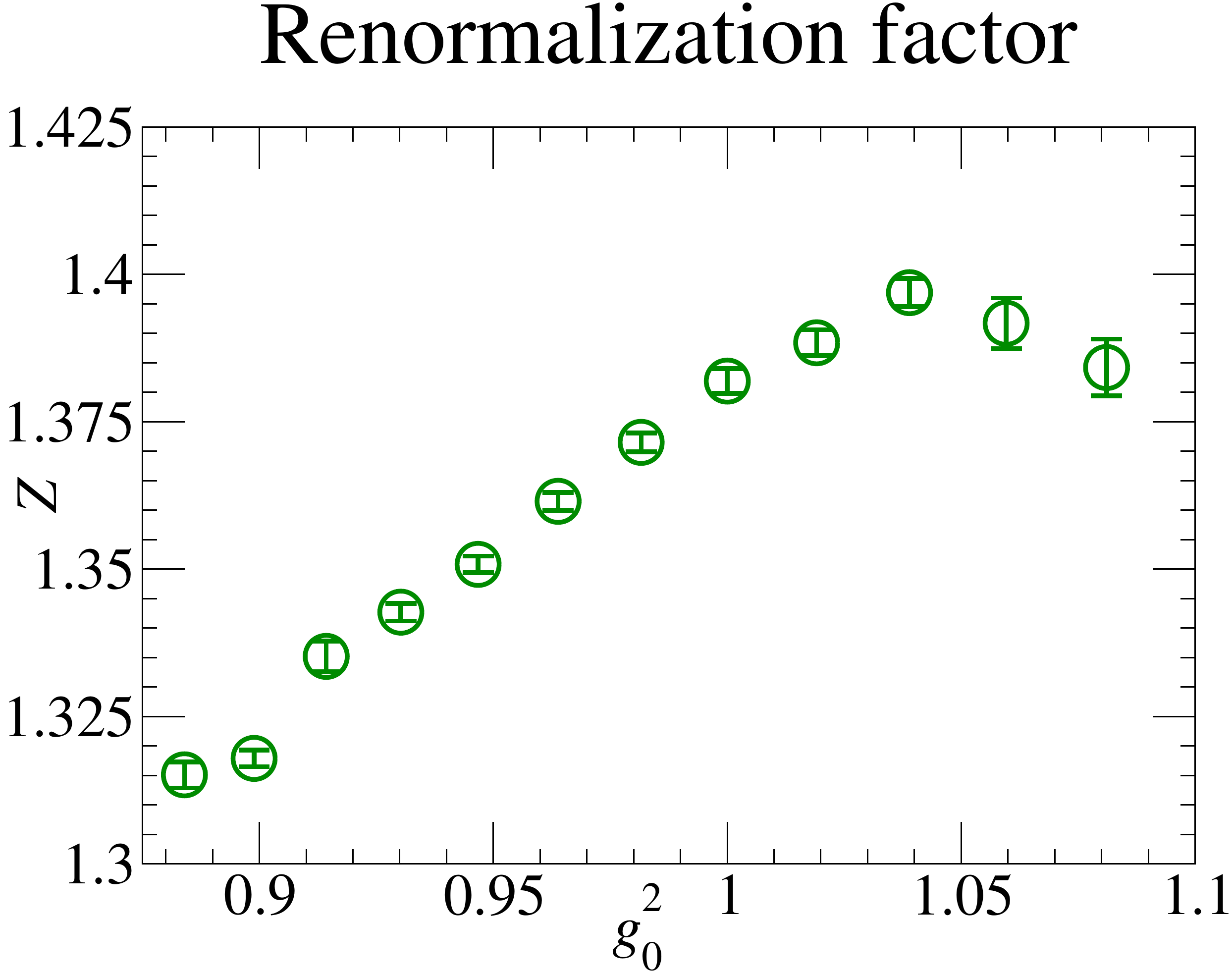}}
  \caption{Setting the scale for simulations of the $\SU(4)$ Yang-Mills theory with the tree-level improved action: the left-hand-side and central plots show, respectively, the zero-temperature string tension in lattice units $\sigma a^2$, the coefficient of the $1/r$ term in the potential (which is compared to the expectation from Bosonic string theory: $\gamma=-\pi/12$~\cite{Luescherterm}), as a function of $\beta$. Finally, the right-hand-side plot shows the values obtained for the renormalization factor for the fundamental representation $Z$ according to eq.~(\protect\ref{eq:Z}), as a function of the square of the bare gauge coupling $g_0^2$.} \label{fig:sigma_gamma_Z_for_SU4}
\end{figure}

In this section we present some preliminary results from our simulations of the $\SU(4)$ gauge theory with the tree-level improved action. Fig.~\ref{fig:sigma_gamma_Z_for_SU4} shows our results from Cornell fits of the $T=0$ potential from smeared Wilson loops: the three panels show, respectively, the string tension in lattice units, the coefficient of the $1/r$ term (which, in the range of $\beta$-values investigated, appears to be approximately compatible with the prediction from the effective Bosonic string model: $\gamma=-\pi/12$~\cite{Luescherterm}), and the renormalization factor for the fundamental representation $Z(g_0^2)$.

\begin{figure}[-t]
  \centerline{\includegraphics[width=.48\textwidth]{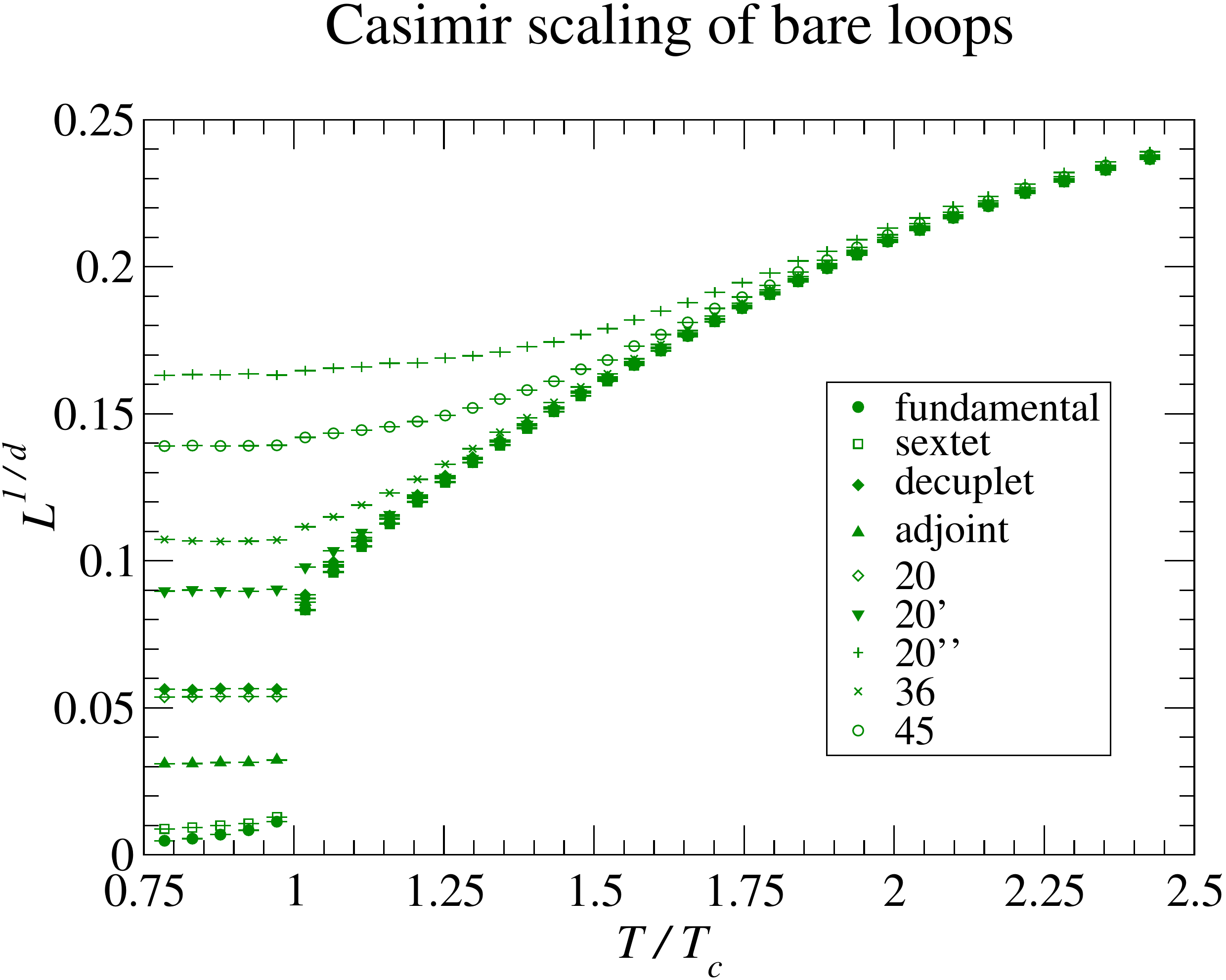} \hspace{4mm} \includegraphics[width=.468\textwidth]{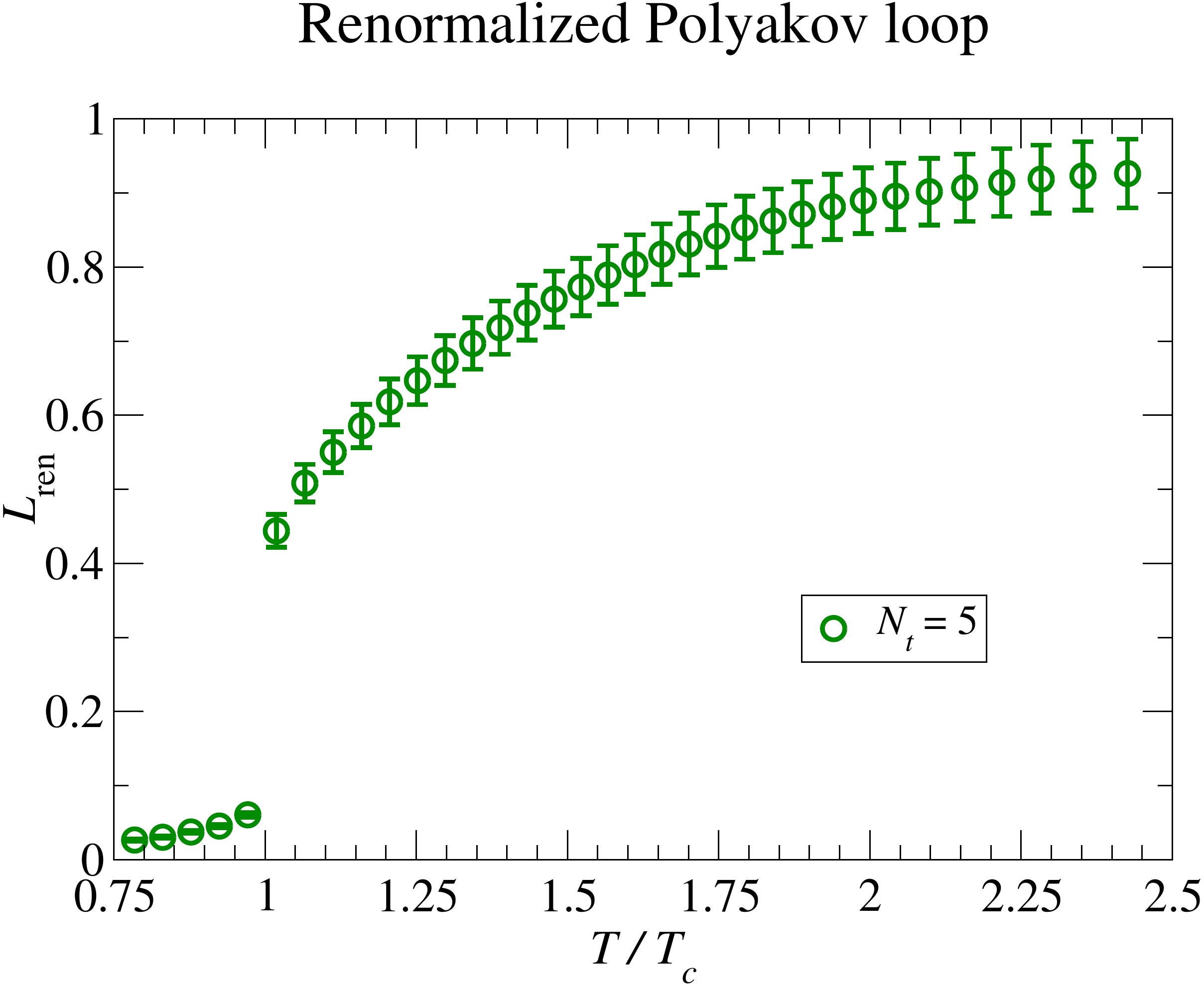}}
  \caption{Left-hand-side panel: Temperature dependence of bare $\SU(4)$ Polyakov loops in different representations, after rescaling their free energies by a factor proportional to the eigenvalue of the corresponding quadratic Casimir $\langle C_2 (r) \rangle $. Right-hand-side panel: The renormalized Polyakov loop in the fundamental representation of $\SU(4)$, as a function of the temperature. The results displayed are obtained from simulations with the improved action, eq.~(\protect\ref{eq:improved_action}).} \label{fig:Casimir_and_renormalized}
\end{figure}

In the left-hand-side panel of fig.~\ref{fig:Casimir_and_renormalized}, we investigate the Casimir scaling in $\SU(4)$ gauge theory: the figure shows the behavior of bare Polyakov loops in different representations, rescaling their free energies according to the ratio of the corresponding eigenvalue of the quadratic Casimir over the one in the fundamental representation: $d=\langle C_2 (r) \rangle / \langle C_2 (f) \rangle$. As one can see, Casimir scaling appears to hold for all representations (up to small deviations, due to finite-volume effects, which affect the highest representations), down to temperatures very close to $T_c$.

Finally, the right-hand-side panel of the same figure shows our preliminary results for the renormalized Polyakov loop in the fundamental representation, as a function of the temperature. These data (obtained on a lattice with $N_t=5$ sites in the Euclidean time direction and $N_s=20$ sites in each of the spacelike directions) show that, similarly to what happens in the $\SU(3)$ Yang-Mills theory~\cite{renormalization_methods}, also in the theory with $\SU(4)$ gauge group the renormalized Polyakov loop jumps to a value close to $1/2$ at $T=T_c$, and then grows up, reaching values near one already at temperatures around $2T_c$. Although from this figure it appears that the renormalized Polyakov loop never goes above the value $1$, we emphasize that these preliminary results at finite lattice spacing, finite volume, and in a limited temperature range do not allow us to draw any conclusion about this issue, and are not in contradiction with the findings obtained in previous studies for $N=3$~\cite{renormalization_methods}.

We are currently extending our calculations to larger and finer lattices, to higher temperatures, and to larger values of $N$.

\acknowledgments

This work is supported by the Academy of Finland, project 1134018. A.M. acknowledges support from the Magnus Ehrnrooth Foundation. Part of the simulations was performed at the Finnish IT Center for Science (CSC), Espoo, Finland. We thank F.~Gliozzi, O.~Kaczmarek, R.~D.~Pisarski, K.~Tuominen and M.~Veps\"al\"ainen for useful discussions.

\end{document}